# Undulator radiation in the THz range

L.A.Gabrielyan, Y.A.Garibyan, Y.R.Nazaryan, K.B.Oganesyan*, M.A.Oganesyan,
M.L.Petrosyan,
Yerevan Physics Institute, Yerevan, Armenia

E.A. Ayryan
Joint Institute for Nuclear Research

*bsk@yerphi.am

**Abstract:** The experimental device for generation of undulator radiation in terahertz wavelength region by use of undulator on ferromagnets is created. The device is based on a beam of a microtrone with the energy $7,5\ MeV$. The radiation wavelength is $200\ \mu$. Registered spontaneous radiation has a power $10^{-6}\ W$ at a current of a beam $2\ mA$ in a pulse. With the optical resonator, in a mode, the amplification of 6 % is received, that in some times is more than expected value. This effect is explained as a result of partial coherence of radiation.

## 1. Introduction

During the last decade there was a sudden increase in the number of the fundamental and applied studies in the field of generation and application of radiation in the wavelength interval of 30 microns up to 0.3 mm that corresponds to the frequency range of 10 - 1 THz [1-35].

We developed and produced the undulator on constant magnets [15]. On the basis of that undulator the experimental installation for generation of spontaneous undulator radiation in terahertz wavelength range was created. The diagram of the installation is shown in Fig. 1.

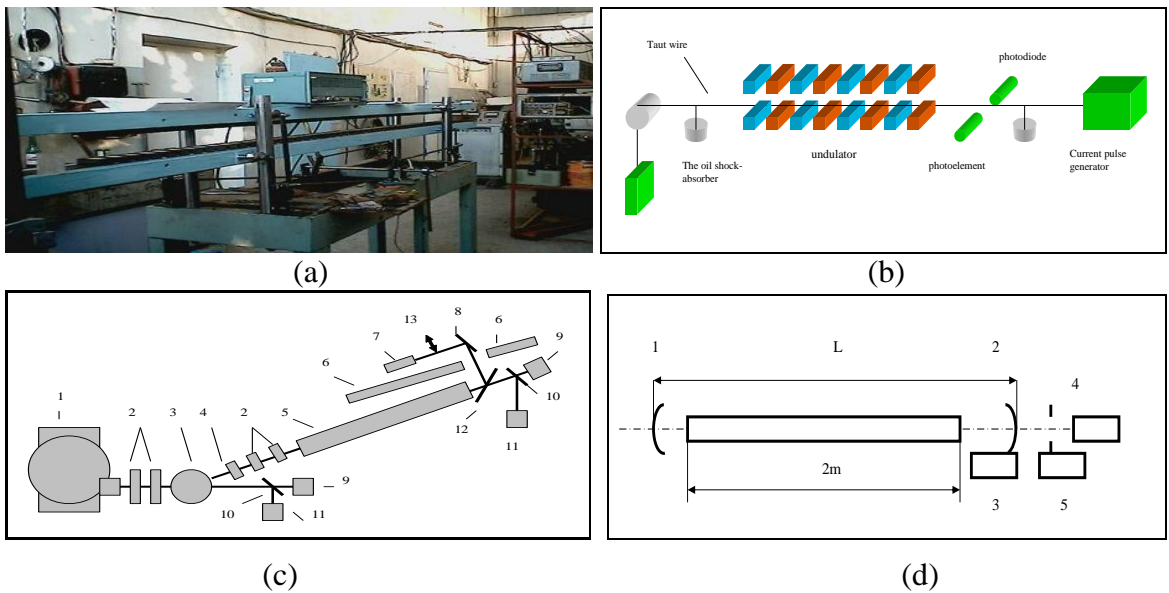

Fig. 1. (a) General view, (b) The scheme of installation for measurement of integrated characteristics of a undulator magnetic field. (c) The diagram of the experimental installation. 1 – microthrone; 2 - quadrupole lenses; 3 - rotating magnet; 4 – correcting magnet; 5 - undulator magnet; 6 - lead protection; 7 – pyrosensor; 8 - spherical mirror; 9 – Faraday cylinder; 10 – luminophor screen. 11 - camera; 12 - thin flat aluminum foil mirror; 13- adjustable blind. (d) The layout of the optical resonator. 1 - spherical mirror with radius of curvature 1.5 m; 2 - output spherical mirror with radius of curvature 1.5 m and with

5 mm diameter of the central opening; 3 - device for remote navigation of the mirror for synchronization of longitudinal modes, 4 – sensor for measurement of the radiation intensity, 5 – distantly operated blind.

## 2. Results

The expected pulse power of the radiation was estimated using [36] formula:

$$P_T(W) = 3.7 \cdot 10^{-8} (\gamma k)^2 N \frac{I(A)}{\lambda_U(m)} \tag{1}$$

where $\kappa = 0.9$ – undulator coefficient, $\gamma = 15$ - Lorentz factor, $N = 22$ - number of the periods, $\lambda_v = 9$ cm - length of the undulator period, $I_A$ - Alfven current.

When microthrone current beam in an impulse equaled 2 mA the expected power in an impulse was $3.6 \cdot 10^{-6}$ W. Using focusing mirror $2.2 \cdot 10^{-5}$ W/cm$^2$ of the radiation intensity in focus was obtained. At the same time the diameter of the beam size was 4 mm and the diameter of the undulator beam track output was 16 mm.

The radiation wavelength was equal to 200 μm. Direct detection in this wavelength range is very complicated. Calorimetric (bolometric, thermocouple) or pyroelectric methods of registration are most accessible. Due to a big inertance, the calorimetric detectors are mainly capable to register only average-power radiation.

Manufacturer-predicted sensitivity of pyrosensor MG-30 used by us is $10^9$ W/(cm$^2$Hz$^{1/2}$). In view of the duration and the frequency of the impulse iteration and cross-section size of the radiation the real sensitivity became $6 \cdot 10^{-7}$ W/(cm$^2$).

The measurements have shown that radiation power is about $2 \cdot 10^{-6}$, i.e. it is 10 times less than expected. It can be explained by the lack of consideration for some losses during the estimation of the expected power of radiation, such as losses through the beam track window, mirrors, pyroelectric sensor window and diffraction losses.

To study the possibility of obtaining the generation in FEL mode the installation was complimented by the optical resonator. Its layout is presented in Fig. 1d.

The full length of the optical resonator (i.e., distance between its mirrors) L is defined from a condition of synchronization of the longitudinal mode $2L = nc/f$, where $c$ - speed of light, $f$ - frequency of the electronic clots passage, $n$ - an integer number. For high frequency microthrone system $f = 2780$ MHz, therefore $L = n \cdot 10.6$ cm. The most accessible length of 3.7 m has been chosen.

In this case FEL works in the impulse mode that limits the opportunity to obtain the most probable power due to the small duration of the beam current impulse.

Let's consider development of the radiation power inside the resonator in view of spontaneous radiation. The amplification coefficient at one passage of radiation G is defined as $P_n/P_{n-1}$, which takes on values $G \geq 1$. At $G = 1$ there is no amplification. Loss coefficient of the optical resonator is defined as $(P_n - \Delta P_n)/P_n$, which takes on values $0 < K < 1$.

The value of power for each passage of radiation impulse through the resonator is:
$P_1 = P_0$
$P_2 = P_0 G\,K + P_0 = P_0 + P_0 G\,K$
$P_3 = P_2 G\,K + P_0 = P_0 (G\,K)^2 + P_0 G\,K + P_0 = P_0 + P_0 G\,K + P_0 (G\,K)^2$
$P_4 = P_3 G\,K + P_0 = P_0 (G\,K)^3 + P_0 (G\,K)^2 + P_0 G\,K + P_0 = P_0 + P_0 G\,K + P_0 (G\,K)^2 + P_0 (G\,K)^3$

$$P_n = P_0 \sum_{0}^{n-1} (KG)^n = G_{FEL} P_0, \tag{2}$$

where $P_0$ is a power of the starting impulse, which in this case is a power of the spontaneous radiation.

The given expression is an infinite geometrical progression. Its limit at the convergence of series is defined by:

$$a_0 + a_0 x + a_0 x^2 + \ldots = \sum_{n=0}^{\infty} a_0 x^n = a_0 \frac{1}{1-x}$$

When G K< 1 series converges, and for the sum we have:

$$P_n = P_0 \frac{1}{1-KG} \qquad (3)$$

or

$$\Delta P_n = P_0,$$

It satisfies the law of conservation of energy, i.e. the power inside the resonator increases to the amount needed for the losses to be equal to the power of refill of the energy.

When GK > 1 the series diverges and power in the optical resonator should be calculated using simple summation.

Calculated values of amplitudes of the power (depending on amplification coefficient, and when the number of passages of radiation in the resonator n = 100 (duration of an impulse of an electronic beam equals 2 µc) at FEL output are shown in Table 1.

Table 1

| GK | 0.95 | 0.96 | 0.97 | 0.98 | 0.99 | 1.0 | 1.01 | 1.02 | 1.03 | 1.04 | 1.05 | 1.1 | 1.15 |
|---|---|---|---|---|---|---|---|---|---|---|---|---|---|
| P/ $P_0$ | 19 | 24 | 31 | 43 | 63 | 100 | 106 | 319 | 626 | 1288 | 2742 | $1,5 \cdot 10^5$ | $9 \cdot 10^6$ |
| $P_{output}$/ $P_0$ = 1% | | | | | | | | | | | | | |
| $P_{output}$/ $P_0$ | 0.19 | 0.24 | 0.31 | 0.43 | 0.63 | 1.0 | 1.06 | 3.19 | 6.26 | 12.88 | 27.42 | $1.5 \cdot 10^4$ | $9 \cdot 10^5$ |
| $P_{output}$/ $P_0$ = 1.5% | | | | | | | | | | | | | |
| $P_{output}$/ $P_0$ | 0.28 | 0.36 | 0.46 | 0.64 | 0.94 | 1.5 | 1.59 | 4.8 | 9.4 | 19.2 | 41.1 | $2.25 \cdot 10^3$ | $1.35 \cdot 10^5$ |
| $P_{output}$/ $P_0$ = 5 % | | | | | | | | | | | | | |
| $P_{output}$/ $P_0$ | 0.96 | 1.2 | 1.55 | 2.15 | 3.15 | 5.0 | 5.3 | 16 | 31 | 64.4 | 137 | $7.5 \cdot 10^3$ | $4.5 \cdot 10^5$ |

As shown in Table 1 in case of amplification factor is more than 0.99 and the output coefficient is 1.5% it is possible to obtain the power of radiation on the output of the installation equal to the power of spontaneous radiation. On the basis of our operational experience [8], it is possible to assume, that losses in the optical resonator are about 5%. Thus, in order to have output power equal to power of the spontaneous radiation amplification coefficient should be 5.8%.

To detect radiation on the output of the optical resonator at output coefficient 1.5% some preliminary measurements were carried out. The results of measurements have shown, that the output signal is on the order of spontaneous radiation, i.e. the amplification coefficient is 6 %.

The amplification factor can be estimated using formula [9]:

$$G_{max} = 1.08 \pi^2 N^3 \lambda_w^2 \frac{i}{I_A \gamma^3 S_m} K^2 f(K), \qquad (5)$$

where $S_m$ is the electron beam section, $f(K) = 0.37$ for flat undulator at K = 1.

When parameters of the beam and undulator are put into formula (5) then G value is equal to 1.1% . Thus measured value of the amplification coefficient is much higher than value obtained using the formula. Such disagreement can be explained by partial coherence of the radiation taking into account that the length of the electron clots is about 3-4 mm and length of the radiation wave is 0.2 mm.

The high degree of the undulator radiation coherence can be achieved when longitudinal $\Delta z = a$ and lateral $\Delta x$; $\Delta y$ dimensions of the clots satisfy following conditions: $\Delta z \leq \lambda$, $\Delta x \approx \Delta y = 2\gamma\lambda$ ($\gamma$ - reduced energy of the particles; $\lambda$ - length of the radiation wave).

Due to the fact that the condition for the lateral sizes is $2\gamma >> 1$ weaker than for the longitudinal sizes the degree of the radiation coherence is defined by the longitudinal size (phase extent) of the clot.

Study [9] is devoted to the radiation of electronic clots of any structure at any movement in homogeneous environments. The common formula for frequency and angular distribution of electron radiation intensity in a clot was obtained. To express an average intensity form-factor is used. It consists of the outcome of two functions, one of which does not depend on the type of the radiation and is defined by longitudinal distribution of electrons (longitudinal form-factor) only, and the other depends on the type of the radiation and lateral distribution (lateral form-factor).

When radiation of the clot was calculated, it was assumed, that all particles were moving with constant speed $v$. In this case, radiation intensity of the clot $I_N$ can be presented in the form of the product of the radiation intensity of separate particle $I_j$ and clot form-factor of $S_N$:

$$\frac{dI_N}{dOd\omega} = \frac{dI_j}{dOd\omega} S_N \qquad (6)$$

$$S_N = N^2 F + N(1-N) \qquad (7)$$

where $F$ - the clot coherence factor

$$F = F_z(\omega) F_\rho(\omega, \theta, \varphi) \qquad (8)$$

$\omega$ - radiation frequency; $\theta$, $\varphi$ - angle ways of the unit vector; Fz (w) and F$_r$ (w, q, j) - longitudinal and lateral coherence factors.

When $F = 0$, the radiation is noncoherent and its intensity is proportional to $N$; at $F = 1$, the radiation is completely coherently and its intensity is proportional to $N^2$. The radiation coherent component is comparable to noncoherent radiation when $F \approx 1/N$. In case of $F >> 1/N$, noncoherent radiation can be neglected.

In the most experimentally feasible cases it is possible to put $\Phi(\omega, \theta, \varphi) = 1$. Then we shall obtain:

$$\left\langle \frac{dI_N}{d\omega} \right\rangle = N^2 \frac{dI_1}{d\omega} F(\omega) \qquad (9)$$

Distribution of electron clots density obtained by standard methods can be well enough approximated by Gauss distribution. However this distribution can be modified in such a manner that approximations of parabolic and Gauss-parabolic functions will appear more convenient.

Let's allow, that function of distribution in the longitudinal direction of electrons is asymmetric, and it can be written down in the form of

$$f(z) = f_{1z}(z)\theta(-z) + f_{2z}(z)\theta(z),$$

$$\theta(z) = 1, \quad z \geq 0$$

$$\theta(z) = 0, \quad z < 0$$

For parabolic distribution of the charge in the clot functions $f_{1z}(z)$ and $f_{2z}(z)$ look like

$$f_{1z}(z) = \frac{3}{2\alpha\gamma_0}\left(1 - z^2 / p^2 z_0^2 \gamma_0^2\right)$$

$$f_{2z}(z) = \frac{3}{2\alpha\gamma_0}\left(1 - z^2 / z_0^2 \gamma_0^2\right) \quad (10)$$

$$\gamma_0 = \left(\frac{\sqrt{e}}{\sqrt{e}-1}\right)^{1/2} \approx 1{,}59$$

where $\gamma_0 a$ - is the maximal longitudinal size of the parabolic clot;
$p$ - defines degree of the particle distribution asymmetry in the longitudinal direction (fig. 3-6-1).

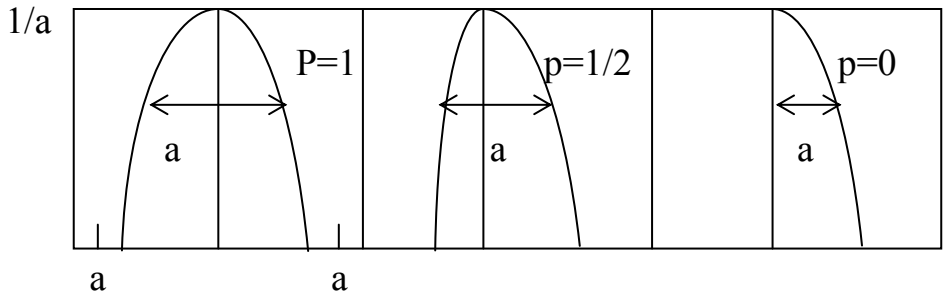

Figure 3-6-1. Distribution of particles in the longitudinal direction.

At p=1 expression for $Fp(\xi)$ has the following description [45]:

$$F_1(\xi) = \left(\frac{3}{x^2}\right)^2 \frac{1}{1+tg^2 x}\left(1 - \frac{tgx}{x}\right)^2 \quad (11)$$

where $\quad x = \gamma_0 \xi / 2, \quad \gamma_0 = \left(\frac{\sqrt{e}}{\sqrt{e}-1}\right)^{1/2} \approx 1{,}5942, \quad \xi = \frac{\omega\alpha}{v} = \frac{2\pi\alpha}{\lambda\beta}$

For values $\xi = 94.25, \quad x = 74.93, \quad N = 1{,}3 \cdot 10^6$ we have $F_1(\xi) = 9 \cdot 10^{-6}$ and $NF \sim 11$.

Thus the partial coherence can lead to the radiation amplification of one order, which agrees with obtained experimental results.

## 3. Conclusion

The generation of undulator radiation in terahertz wavelength region is investigated. The radiation wavelength is 200 $\mu$. Registered spontaneous radiation has a power $10^{-6}$ W at a current of a beam 2 $mA$ in a pulse. With the optical resonator, the amplification of 6 % is received, that in some times is more than expected value. This effect is explained as a result of partial coherence of radiation.